\documentstyle[12pt]{article}
\catcode`\@=11
\long\def\@makefntext#1{
\protect\noindent \hbox to 3.2pt {\hskip-.9pt
$^{{\ninerm\@thefnmark}}$\hfil}#1\hfill}		%CAN BE USED

\def\@makefnmark{\hbox to 0pt{$^{\@thefnmark}$\hss}}  %ORIGINAL
	
\def\ps@myheadings{\let\@mkboth\@gobbletwo
\def\@oddhead{\hbox{}
\rightmark\hfil\ninerm\thepage} 
\def\@oddfoot{}\def\@evenhead{\ninerm\thepage\hfil
\leftmark\hbox{}}\def\@evenfoot{}
\def\sectionmark##1{}\def\subsectionmark##1{}}

\renewcommand{\thefootnote}{\fnsymbol{footnote}}

\newcounter{sectionc}\newcounter{subsectionc}\newcounter{subsubsectionc}
\renewcommand{\section}[1] {\vspace*{0.6cm}\addtocounter{sectionc}{1}
\setcounter{subsectionc}{0}\setcounter{subsubsectionc}{0}\noindent
	{\normalsize\bf\thesectionc. #1}\par\vspace*{0.4cm}}
\renewcommand{\subsection}[1] {\vspace*{0.6cm}\addtocounter{subsectionc}{1}
	\setcounter{subsubsectionc}{0}\noindent
	{\normalsize\it\thesectionc.\thesubsectionc. #1}\par\vspace*{0.4cm}}
\renewcommand{\subsubsection}[1]
{\vspace*{0.6cm}\addtocounter{subsubsectionc}{1}
	\noindent {\normalsize\rm\thesectionc.\thesubsectionc.\thesubsubsectionc.
	#1}\par\vspace*{0.4cm}}

\newcounter{appendixc}
\newcounter{subappendixc}[appendixc]
\newcounter{subsubappendixc}[subappendixc]

\renewcommand{\appendix}[1] {\vspace*{0.6cm}
        \refstepcounter{appendixc}
        \setcounter{figure}{0}
        \setcounter{table}{0}
        \setcounter{equation}{0}
        \renewcommand{\thefigure}{\Alph{appendixc}.\arabic{figure}}
        \renewcommand{\thetable}{\Alph{appendixc}.\arabic{table}}
        \renewcommand{\theappendixc}{\Alph{appendixc}}
        \renewcommand{\theequation}{\Alph{appendixc}.\arabic{equation}}
        \noindent{\bf Appendix \theappendixc #1}\par\vspace*{0.4cm}}

\def\abstracts#1{{
	\centering{\begin{minipage}{12.2truecm}\footnotesize\baselineskip=12pt\noindent
	\centerline{\footnotesize ABSTRACT}\vspace*{0.3cm}
	\parindent=0pt #1
	\end{minipage}}\par}}

\renewenvironment{thebibliography}[1]
	{\begin{list}{\arabic{enumi}.}
	{\usecounter{enumi}\setlength{\parsep}{0pt}
\setlength{\leftmargin 1.25cm}{\rightmargin 0pt}
	 \setlength{\itemsep}{0pt} \settowidth
	{\labelwidth}{#1.}\sloppy}}{\end{list}}

\newcounter{itemlistc}
\newcounter{romanlistc}
\newcounter{alphlistc}
\newcounter{arabiclistc}

\newcommand{\fcaption}[1]{
        \refstepcounter{figure}
        \setbox\@tempboxa = \hbox{\footnotesize Fig.~\thefigure. #1}
        \ifdim \wd\@tempboxa > 6in
           {\begin{center}
        \parbox{6in}{\footnotesize\baselineskip=12pt Fig.~\thefigure. #1}
            \end{center}}
        \else
             {\begin{center}
             {\footnotesize Fig.~\thefigure. #1}
              \end{center}}
        \fi}

\newcommand{\tcaption}[1]{
        \refstepcounter{table}
        \setbox\@tempboxa = \hbox{\footnotesize Table~\thetable. #1}
        \ifdim \wd\@tempboxa > 6in
           {\begin{center}
        \parbox{6in}{\footnotesize\baselineskip=12pt Table~\thetable. #1}
            \end{center}}
        \else
             {\begin{center}
             {\footnotesize Table~\thetable. #1}
              \end{center}}
        \fi}

\def\@citex[#1]#2{\if@filesw\immediate\write\@auxout
	{\string\citation{#2}}\fi
\def\@citea{}\@cite{\@for\@citeb:=#2\do
	{\@citea\def\@citea{,}\@ifundefined
	{b@\@citeb}{{\bf ?}\@warning
	{Citation `\@citeb' on page \thepage \space undefined}}
	{\csname b@\@citeb\endcsname}}}{#1}}

\newif\if@cghi
\def\cite{\@cghitrue\@ifnextchar [{\@tempswatrue
	\@citex}{\@tempswafalse\@citex[]}}
\def\citelow{\@cghifalse\@ifnextchar [{\@tempswatrue
	\@citex}{\@tempswafalse\@citex[]}}
\def\@cite#1#2{{$\null^{#1}$\if@tempswa\typeout
	{IJCGA warning: optional citation argument
	ignored: `#2'} \fi}}

 1
 1
 1

\font\ninerm=cmr9

\newcommand{\be}{\begin{equation}}
\newcommand{\ee} {\end{equation}}
\newcommand{\bea}{\be\begin{array}}
\newcommand{\eea} {\end{array}\ee}
\begin{document}
\begin{flushright}
{ALBERTA THY. 31-95\\ December, 1995}
\end{flushright}
%\centerline{\normalsize\bf WORLD SCIENTIFIC PUBLISHING COMPANY}
\baselineskip=22pt
\centerline{\normalsize\bf Nonfactorization in B and D decays}
\baselineskip=16pt
%\centerline{\normalsize\bf MANUSCRIPT BY COMPUTER}
%\centerline{\footnotesize\sf (For subsequent 20\% photoreduction
%to 17.8 $\times$ 11.9 cm text area)

%\vfill
%\vspace*{0.6cm}
\centerline{\footnotesize A. N. Kamal}
\baselineskip=13pt
\centerline{\footnotesize\it  Department of Physics, University of  Alberta,
}
\baselineskip=12pt
\centerline{\footnotesize\it  Edmonton, AB.  T6G 2J1, Canada}
\centerline{\footnotesize E-mail: kamal@phys.ualberta.ca}
%\vspace*{0.3cm}
%\centerline{\footnotesize and}
%\vspace*{0.3cm}
%\centerline{\footnotesize SECOND AUTHOR'S NAME}
%\baselineskip=13pt

%\vfill
\vspace*{0.9cm}
\abstracts{ We discuss the role of nonfactorized contributions in $B \rightarrow \psi + {K}^{*} $ and charmed meson decays. We demonstrate, using $ {D}^{+}_{S} \rightarrow \phi \pi$ as  a model, how nonfactorization, annihilation and final state interactions can be built into effective and unitarized $ {a}_{1} $ and $ {a}_{2} $.}

%\vspace*{0.6cm}
\normalsize\baselineskip=15pt
\setcounter{footnote}{0}
\renewcommand{\thefootnote}{\alph{footnote}}

\section{Introduction}
Factorization approximation, where the matrix element of a product of two  color-singlet weak currents is approximated by the product of the matrix elements of the individual currents, is the most commonly used scheme to calculate the  amplitudes for two-body hadronic decays of   B  and D mesons  \cite{1}. It was shown recently  \cite{2}  that this  approximation failed to reproduce the longitudinal polarization observed in $B \rightarrow ~\psi ~+ ~{K}^{*} $ decay in all the commonly used models of form factors . Subsequently it was shown \cite{3}  how a small nonfactorization  contribution helps in understanding the longitudinal polarization in   $B \rightarrow \psi + {K}^{*} $ decay in all the commonly used models of form factors .

In the following I will discuss three topics :
(I) Nonfactorizatation and polarization in color-suppressed  $ B \rightarrow \psi + {K}^{*} $ decay,
(II) Nonfactorization  in Cabibbo-favored   D decays  and
(III) Nature of ${a}_{1}$and $ {a}_{2} $.

\section{ $B \rightarrow \psi + {K}^{*} $ decay}
The effective weak Hamiltonian  for decays of kind $ b \rightarrow s c \bar{c}$ is :

$$
{{H}_{W}}^{eff} = {{G}_{F} \over \sqrt{2} } {V}_{cb} {V}_{cs}^{*} \left\{  {C}_{1} ( \bar{c} b ) ( \bar{s} c ) + {C}_{2} ( \bar{c} c ) ( \bar{s} b ) \right\}\eqno(1)
$$
where $  ( \bar{c} b )$ etc. represents color-singlet (V-A) current,$ {V}_{cb}$ and $  {V}_{cs}$ are the relevant Cabibbo-Kobayashi-Maskawa (CKM) mixing  parameters and $ {C}_{1} $ and $ {C}_{2} $ are the  QCD  coefficients for which we  use the values \cite{3}:

$$
{C}_{1}=1.12 \pm 0.01, ~~~ {C}_{2}=-0.27 \pm 0.03 \eqno(2)
$$
If we Fierz-transform in color space, we can write
$$
(\bar{c} c) ( \bar{s} b ) = {1 \over {N}_{c}} (\bar{c} b) ( \bar{s} c ) + {1 \over 2} \sum_{a}{ (\bar{c} {\lambda}^{a} b)  ( \bar{s}  {\lambda}^{a} c ) } \eqno(3a)
$$
and
$$
(\bar{c} b) ( \bar{s} c ) = {1 \over {N}_{c}} (\bar{c} c) ( \bar{s} b ) + {1 \over 2} \sum_{a}{ (\bar{c} {\lambda}^{a} c)  ( \bar{s}  {\lambda}^{a} b )} \eqno(3b)
$$
where $ {\lambda}^{a}$ are the Gell-Mann matrices and $ {N}_{c}= 3 $ is the number of colors.It is convenient to define two parameters ${a}_{1}$ and $ {a}_{2} $ as follows:
$$
{a}_{1} = {C}_{1} + {1 \over {N}_{c}} {C}_{2}= 1.03 \pm 0.01\eqno(4a)
$$
$$
{a}_{2} = {C}_{2} + {1 \over {N}_{c}} {C}_{1}= 0.10 \pm 0.03 \eqno(4b)
$$

The decay amplitude for $ B \rightarrow \psi + {K}^{*} $ can be written in the following  form :

$$
A (B \rightarrow \psi + {K}^{*} ) = {{G}_{F} \over \sqrt{2} } {V}_{cb} {V}_{cs}^{*} \left\{ {a}_{2} < {K}^{*}  \psi \mid  ( \bar{c} c ) ( \bar{s} b )\mid B > + {C}_{1} < {K}^{*}  \psi \mid {{\tilde{H}}_{W}}^{(8)}\mid B >  \right\}\eqno(5)
$$
where
$$
{{\tilde{H}}_{W}}^{(8)}= {1 \over 2} \sum_{a}{ (\bar{c} {\lambda}^{a} c)  ( \bar{s}  {\lambda}^{a} b )}\eqno(6)
$$
is a product of color-octet currents. In the factorization approximation, only the first term in Eq.(5) is kept.

Confining ourselves to the factorization approximation, we can write the following expressions for the decay amplitude in terms of  the relevant form factors \cite{1} :
\begin{eqnarray}
A (B \rightarrow \psi + {K}^{*} ) = & \tilde{{G}}_{F} {a}_{2} {m}_{\psi}{f}_{\psi} & \left\{ \frac{}{}\left( {m}_{B} + {m}_{{K}^{*}} \right) {{A}_{1}}^{B{K}^{*}} ( {m}_{\psi}^{2} ) \right. \nonumber \\
& &~  - \left. {2\over  \left( {m}_{B} + {m}_{{K}^{*}} \right)} {\epsilon}_{1}\cdot {p}_{B} {\epsilon}_{2}\cdot {p}_{B} {{A}_{2}}^{B{K}^{*}} ( {m}_{\psi}^{2} )\right .\nonumber \\
&  &~+\left. {2i \over  \left( {m}_{B} + {m}_{{K}^{*}} \right)} {\epsilon}_{\mu \upsilon \rho \sigma}
 {{\epsilon}_{1}}^{\mu} {{\epsilon}_{2}}^{\nu} {{p}_{{K}^{*}}}^{\rho} {{p}_{B}}^{\sigma} {V}^{B{K}^{*}} ( {m}_{\psi}^{2}) \right\} ~(7) \nonumber
\end{eqnarray}
where $ \tilde{{G}}_{F} =  {{G}_{F} \over \sqrt{2} } {V}_{cb} {V}_{cs}^{*}$ and $ {\epsilon}_{1}$
and $ {\epsilon}_{2}$ are the polarization vector of the $ \psi$ and $ {K}^{*}$ respectively.From Eq.(7) one obtains \cite{2}
$$
B ( B   \rightarrow \psi + {K}^{*} ) = 2.84 {{a}_{2}}^{2} \mid  {{A}_{1}}^{B{K}^{*}} ( {m}_{\psi}^{2} ){\mid}^{2} ( \Sigma_{LL} +  \Sigma_{TT} )   \% \eqno(8)
$$
where  for longitudinal and transverse states of polarization,

$$
\Sigma_{LL}= {(a -bx)}^{2},~~~ \Sigma_{TT}= 2 ( 1 + {c}^{2} {y}^{2} ) \eqno(9)
$$
with

$$
x\equiv {{A}_{2}( {m}_{\psi}^{2}) \over  {A}_{1}( {m}_{\psi}^{2})}~,~~~~~~~ y\equiv {V( {m}_{\psi}^{2} )\over {A}_{1}( {m}_{\psi}^{2})}\eqno(10a)
$$
and
$$
 a=3.147,~ b=1.297, ~c=0.434 \eqno(10b)
$$

From Eq.(9), we obtain the longitudinal polarization as follows:

$$
{P}_{L} = { \Sigma_{LL}\over  \Sigma_{LL} +  \Sigma_{TT}} = {{(a -bx)}^{2} \over  {(a -bx)}^{2} +  2 ( 1 + {c}^{2} {y}^{2} ) }\eqno(11)
$$
Experimentally \cite{4,5}
$$
{P}_{L} = 0.78 \pm 0.07 .  \eqno(12)
$$

In Figure 1 we have shown  the allowed domain in (x-y) plane to one standard  deviation of the central value of ${P}_{L}$ data.  We have also shown the value of  x and y predicated by six different models. For details the reader is referred to Ref. [2].
\begin{figure}
\begin{picture}(430,200)(0,0)
%\put(430,0){-}
%\put(430,260){-}
%\put(0,0){-}
%\put(0,260){-}
\includegraphics{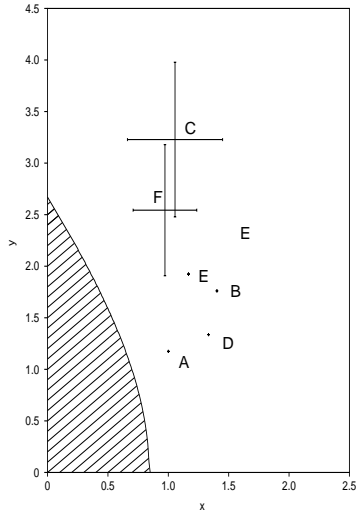}
\end{picture}
\caption{The domain of x and y allowed by the polarization data Eq.(12). Points A,B, \ldots represent predicted values of x and y in various models. See Ref. [2] for details. } \label{}
\end{figure}

It is clear  that the commonly used models of form factors generate points in (x-y)  plane that are several standard deviations removed from those required by polarization data.

Assume now that there are nonfactorized contributions to the decay amplitude of Eq. (5).These could arise from two sources :    (i) nonfactorized contribution to the first term comprised of color-singlet  currents and (ii) contribution from $ {\tilde{{H}_{w}}}^{(8)}$ which is a product of two color-octet currents.  This latter contribution is enhanced  relative to the former  by the fact that $ \mid {C}_{1}\mid \approx 10 \mid {a}_{2}\mid$. If for simplicity  we assign   nonfactorized contribution to the Lorentz structure  belonging to $ {A}_{1}$ only, then we need to modify our formalism with the replacement
$$
{{A}_{1}}^{B{K}^{*}}  ( {m}_{\psi}^{2} )  \rightarrow  {{A}_{1}}^{B{K}^{*}}  ( {m}_{\psi}^{2} ) + {{A}_{1}}^{(1)nf} + {{C}_{1} \over {a}_{2}}{{A}_{1}}^{(8)nf}\eqno(13)
$$
where ${{A}_{1}}^{(1)nf}  $ and ${{A}_{1}}^{(8)nf} $ are the two nonfactorized contributions referred to  above.

The resulting formula for $ {P}_{L}$ becomes
$$
 {P}_{L} = {{(a \xi - bx )}^{2} \over {\left( a\xi - b x  \right)}^{2} + 2 \left( {\xi}^{2}  + {c}^{2} {y}^{2}\right)} \eqno(14)
$$
where
$$
\xi = 1 + {{C}_{1} \over {a}_{2}} \chi \eqno(15a)
$$
and
$$
 \chi=\left(  {{A}_{1}}^{(8)nf} + {{a}_{2} \over {C}_{1}} {{A}_{1}}^{(1)nf} \right)  / {{A}_{1}}^{B{K}^{*}}  ( {m}_{\psi}^{2} ) \eqno(15b)
$$

In Figure  2 we have plotted $ {P}_{L}$ for different values of  $ \chi$ which is a measure of nonfactorized contribution .
\begin{figure}
\begin{picture}(430,200)(0,0)
%\put(430,0){-}
%\put(430,260){-}
%\put(0,0){-}
%\put(0,260){-}
\includegraphics{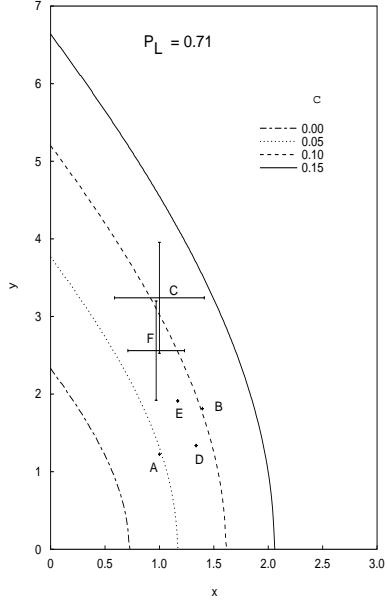}
\end{picture}
\caption{Domains of x and y allowed by polarization data for different values of the nonfactorization parameter $\chi $.} \label{}
\end{figure}

 We note that for $ \chi \approx 0.1 $ all models of form factors accomodate data. We remark that if nonfactorization were put in, say $ {A}_{2} $ , we would have needed much larger values of the parameter analogous to $\chi$. Thus even as little as a $10 \%$ nonfactorized contribution provides an understanding of the longitudinal polarization in all commonly used models of form factors .

\section { Nonfactorization in Cabibbo-favored $  D$  decays. }
In the following we will  discuss the role of nonfactorization in a selected few Cabibbo-favored  $D$ decays: $ D \rightarrow \pi \bar{K}, \pi {\bar{K}}^{*}, \rho \bar{K}$.

The relevant effective Hamiltonian is
$$
{{H}_{W}}^{eff} = {{G}_{F} \over \sqrt{2} } {V}_{cs} {V}_{ud}^{*} \left\{  {C}_{1} ( \bar{u} d ) ( \bar{s} c ) + {C}_{2} ( \bar{u} c ) ( \bar{s} d ) \right\} \eqno(16)
$$
where for $ {C}_{1} $ and $ {C}_{2}$ at charm mass scale we adopt the following values  \cite{6,7} ,
$$
{C}_{1} = 1.26 \pm 0.04, ~~~ {C}_{2}= - 0.51 \pm 0.05 \eqno(17)
$$
so that  with $ {N}_{c}= 3 $, we obtain 
$$
{a}_{1} = 1.09 \pm 0.04, ~~~ {a}_{2} = - 0.09 \pm 0.05 \eqno(18)
$$
Fierz transformation of the current products in color-space leads to
$$
(\bar{u} c) ( \bar{s} d ) = {1 \over {N}_{c}} (\bar{u} d) ( \bar{s} c ) + {1 \over 2} \sum_{a}{ (\bar{u} {\lambda}^{a} d)  ( \bar{s}  {\lambda}^{a} c ) } \eqno(19a)
$$
and
$$
(\bar{u} d) ( \bar{s} c ) = {1 \over {N}_{c}} (\bar{u} c) ( \bar{s} d ) + {1 \over 2} \sum_{a}{ (\bar{u} {\lambda}^{a} c)  ( \bar{s}  {\lambda}^{a} d )} \eqno(19b)
$$
Define
$$
{{H}_{W}}^{(8)}= {1 \over 2}\sum_{a}{ (\bar{u} {\lambda}^{a} d)  ( \bar{s}  {\lambda}^{a} c ) } \eqno(20a)
$$
and
$$
\tilde{H}_{W}^{(8)} = {1 \over 2} \sum_{a}{ (\bar{u} {\lambda}^{a} c)  ( \bar{s}  {\lambda}^{a} d )}\eqno(20b)
$$
Using the above equations we calculate the decay amplitude for  $ D \rightarrow  \bar{K} \pi,  {\bar{K}}^{*} \pi$ and $ \bar{K}   \rho$ decays. As  an example,

$$
A( {D}^{0} \rightarrow {K}^{-} {\pi}^{+} ) =  {{G}_{F} \over \sqrt{2} } {V}_{cs} {V}_{ud}^{*} {f}_{\pi} ( {m}_{D}^{2} - {m}_{K}^{2} ) {{a}_{1}}^{eff} {{F}_{0}}^{DK} ( {m}_{\pi}^{2} )  \eqno(21a)
$$
and
$$
A( {D}^{0} \rightarrow \bar{{K}^{0}} {\pi}^{0} ) =  {{G}_{F} \over \sqrt{2} } {V}_{cs} {V}_{ud}^{*} {f}_{K} ( {m}_{D}^{2} - {m}_{\pi}^{2} ) {{a}_{2}}^{eff} {{F}_{0}}^{D \pi} ( {m}_{K}^{2} ) \eqno(21b)
$$
where

$$
{{a}_{1}}^{eff}= {a}_{1} \left\{ 1 + { {{F}_{0}}^{(1)nf} \over {{F}_{0}}^{DK} ( {m}_{\pi}^{2} ) } + {{C}_{2} \over {a}_{1}}  { {{F}_{0}}^{(8)nf} \over {{F}_{0}}^{DK} ( {m}_{\pi}^{2} ) }\right\} \eqno(22a)
$$
and
$$
{{a}_{2}}^{eff}= {a}_{2} \left\{ 1 + {{\tilde{{F}_{0}}}^{(1) nf }\over {{F}_{0}}^{D\pi} ( {m}_{K}^{2} ) } + {{C}_{1} \over {a}_{2}}  {{\tilde{{F}_{0}}}^{(8) nf } \over {{F}_{0}}^{D\pi} ( {m}_{K}^{2} ) }\right\} \eqno(22b)
$$
$ {{F}_{0}}^{(1)nf} $ and $ {\tilde{{F}_{0}}}^{(1) nf }$ are the nonfactorized contributions  from the product of color-singlet currents and $ {{F}_{0}}^{(8)nf}  $ and $ {\tilde{{F}_{0}}}^{(8) nf} $ arise from $ {{H}_{W}}^{(8)}$ and ${\tilde{{H}_{W}}}^{(8)}$ of Eq. (20),respectively.

>From Eq. (21) we calculate the isospin amplitudes \cite{7,8}, $ {A}_{{1 \over 2}} $ and $ {A}_{{3 \over 2}} $, by setting the phases equal to zero in the following isospin decomposition,

$$
A( {D}^{0} \rightarrow {K}^{-} {\pi}^{+} ) = {1 \over \sqrt{3} } \left\{  {A}_{{3 \over 2}} {e}^{i {\delta}_{{3 \over 2}}} +\sqrt{2}   {A}_{{1\over 2}} {e}^{i {\delta}_{{1 \over 2}}}\right\}
\eqno(23a)
$$
and
$$
 A( {D}^{0} \rightarrow \bar{{K}^{0}} {\pi}^{0} )= {1 \over \sqrt{3} } \left\{ \sqrt{2}  {A}_{{3 \over 2}} {e}^{i {\delta}_{{3 \over 2}}} -   {A}_{{1\over 2}} {e}^{i {\delta}_{{1 \over 2}}}   \right\} \eqno(23b)
$$
Having thus determined $ {A}_{{1 \over 2}} $ and  $ {A}_{{3 \over 2}} $, we reinstated the phases with \cite{8}   $ {\delta}_{{1 \over 2}} -  {\delta}_{{3 \over 2}} = { ( 86 \pm 8 )}^{\circ } $ and calculate the branching ratio. Fitting the branching ratio data determined  $ {{a}_{1}}^{eff}$  and $ {{a}_{2}}^{eff}$ to lie in the following ranges \cite{7},

$$
1.13 \leq  {{a}_{1}}^{eff} \leq 1.17, ~~~ -0.46 \leq {{a}_{2}}^{eff} \leq -0.42 \eqno(24)
$$

We repeated this procedure for $ D \rightarrow {\bar{K}}^{*} \pi $ and $ \bar{K} \rho $ decays where  the difference of isospin phases is known \cite{8}and obtained\cite{7},
$$
{\bar{K}}^{*} \pi : ~~~ 1.74 \leq   {{a}_{1}}^{eff} \leq   1.96,~~~~ -0.53 \leq  {{a}_{2}}^{eff} \leq -0.43
\eqno(25a)
$$
and
$$
\bar{K} \rho : ~~~1.17 \leq   {{a}_{1}}^{eff}  \leq 1.32,~~~~ -1.00 \leq  {{a}_{2}}^{eff} \leq -0.75
\eqno(25b)
$$

In the above calculation we used the measured form factors  as much as possible; ${{F}_{0}}^{DK} ( 0), {{A}_{1}}^{D {K}^{*}}(0), {{A}_{2}}^{D {K}^{*}}(0)$ and $  {V}^{D {K}^{*}}(0)$  from Ref. [10], $ {{F}_{0}}^{D\pi} ( 0)$ from Ref. [8] and the as-yet-unmeasured $ {{A}_{0}}^{D \rho} (0)$ from the model of  Ref. [1]. From Eq. (24) we note that the effective $ {a}_{1}$ and $ {a}_{2} $ are very close to the values advocated in Ref. [1]   which gave rise to the belief that $ {N}_{C}  \rightarrow \infty $  limit was relevant to $ D$ decays. However the value of $  {{a}_{1}}^{eff}$ and $  {{a}_{2}}^{eff}$ determined from $ {\bar{K}}^{*} \pi$ and $ \bar{K} \rho$ decays belie that belief.

\section{ Nature of   ${{a}_{1}}^{eff}$ and $  {{a}_{2}}^{eff} $}

Up to this stage our discussion has assumed that the effective ${{a}_{1}}^{eff}$ and $  {{a}_{2}}^{eff}$  are real .
Here we discuss how final-state interactions can be introduced and how they render  ${{a}_{1}}^{eff}$ and $  {{a}_{2}}^{eff}$  complex . 

To illustrate the ideas, we look at the decay $ {{D}_{S}}^{+}\rightarrow \phi {\pi}^{+}$.It  involves a single isospin amplitude, $ I=1$, in the final state. Before one introduces inelastic final state interactions, the decay amplitude for ${{D}_{S}}^{+}\rightarrow \phi {\pi}^{+}$ is given by
$$
A({{D}_{S}}^{+}\rightarrow \phi {\pi}^{+}) = {{G}_{F} \over \sqrt{2} } {V}_{cs} {V}_{ud}^{*} {{a}_{1}}^{eff} ( 2 {m}_{\phi} ) {f}_{\pi} \epsilon \cdot {p}_{{D}_{S}} {{A}_{0}}^{{D}_{S}\phi} ( {m}_{\pi}^{2} ) \eqno(26)
$$
where
$$
{{a}_{1}}^{eff} = {a}_{1} \left\{ 1 + {{{A}_{0}}^{ (1) nf} \over  {{A}_{0}}^{{D}_{S}\phi} ( {m}_{\pi}^{2} ) }  + {{C}_{2} \over  {a}_{1}} {{{A}_{0}}^{ (8) nf} \over  {{A}_{0}}^{{D}_{S}\phi} ( {m}_{\pi}^{2} ) }  +{{f}_{{D}_{S}} \over  {f}_{\pi}} {{{A}_{0}}^{ ann} \over  {{A}_{0}}^{{D}_{S}\phi} ( {m}_{\pi}^{2} ) } \right\}\eqno(27)
$$
where apart from the nonfactorized contributions  we have also absorbed a real annihilation term in the defenition of an $ {{a}_{1}}^{eff}$  .

We next unitarize the amplitude through final-state interactions . This renders the amplitude complex and  one can define a complex, unitarized $ {{a}_{1}}^{eff}$ as follows (~the superscript $U$ stands for 'unitarized '):

$$
{A}^{U}( {{D}_{S}}^{+}\rightarrow \phi {\pi}^{+})=  {{G}_{F} \over \sqrt{2} } {V}_{cs} {V}_{ud}^{*} {{a}_{1}}^{U, eff} ( 2 {m}_{\phi} ) {f}_{\pi} \epsilon \cdot {p}_{{D}_{S}} {{A}_{0}}^{{D}_{S}\phi} ( {m}_{\pi}^{2} ) \eqno(28)
$$

For the sake of illustration, we consider coupling of the $ \phi {\pi}^{+} $ channel to the G-even $ {K}^{*} K $ channel given by the symmetric state

$$
\mid  {K}^{*} K  {>}_{S} = {1 \over \sqrt{2} }  \left\{  \mid  {{K}^{*} } ^{+}  \bar{{K}^{0}}  {>} +  \mid   {K}^{+} {\bar{{K}^{0}}}^{*} {>}\right\}\eqno(29)
$$
In Figure  3 we have shown schematically how the channels are coupled.
\begin{figure}
\begin{picture}(430,200)(0,0)
%\put(430,0){-}
%\put(430,260){-}
%\put(0,0){-}
%\put(0,260){-}
\includegraphics{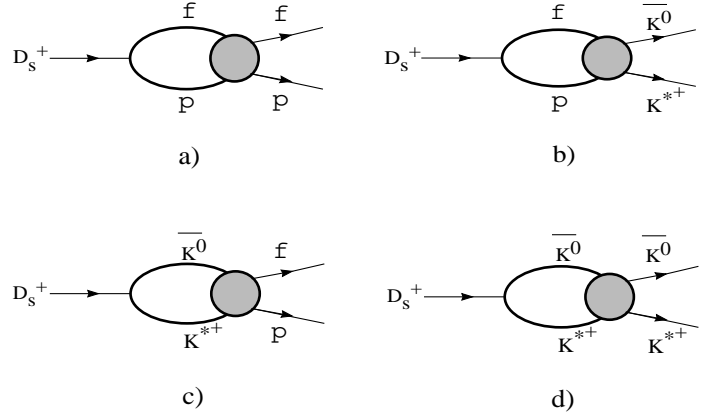}
\end{picture}
\caption{Inelastic final-state interaction Feynman diagrams} \label{fiberbundle}
\end{figure}

We adopt the K-matrix formalism \cite{11} to unitarize the decay amplitude. The unitarized decay amplitude, are given in terms of the following coupled equations,

$$
\left (\matrix{
{A}^{U, \phi \pi}\cr
{A}^{U, ( {K}^{*} K {)}_{S} }\cr
}\right ) = {\left( {\bf 1} - i {{\bf k}}^{3} {\bf K} \right)}^{-1 T }
\left (\matrix{
{A}^{\phi \pi}\cr
{A}^{ ( {K}^{*} K {)}_{S} }\cr
}\right )\eqno(30)
$$
where $ {\bf K} $ is a real, symmetric  $2\times 2 $ matrix with dimensions $ {( GeV)}^{-3}  , {\bf k}^{3} $ is a diagonal matrix with entries $ { k}_{1}^{3} $ and $ { k}_{2}^{3} $ as appropriate   P-wave  threshold factors for the two channels and $ {A}^{\phi \pi}$ and $ {A}^{ ( {K}^{*} K {)}_{S} }$ are the un-unitarized amplitudes. The latter is constructed out of the following two amplitudes,

$$
A ( {{D}_{S}}^{+}\rightarrow  \bar{{K}^{0}}  {{K}^{*} } ^{+} ) = {{G}_{F} \over \sqrt{2} } {V}_{cs} {V}_{ud}^{*} {{a}_{2}}^{ eff}   {f}_{K} ( 2 {m}_{{K}^{*}} ) \epsilon \cdot {p}_{{D}_{S}} {{A}_{0}}^{{D}_{S}{K}^{*}} ( {m}_{K}^{2} ) \eqno(31)
$$
with
$$
{{a}_{2}}^{eff} = {a}_{2} \left\{ 1 + {{{B}_{0}}^{ (1) nf} \over  {{A}_{0}}^{{D}_{S}{K}^{*}} ( {m}_{K}^{2} ) }  + {{C}_{1} \over  {a}_{2}} {{{B}_{0}}^{ (8) nf} \over  {{A}_{0}}^{{D}_{S}{K}^{*}} ( {m}_{K}^{2} ) }    + {{a}_{1} \over {a}_{2}}{{f}_{{D}_{S}} \over  {f}_{K}} {{{B}_{0}}^{ ann} \over   {{A}_{0}}^{{D}_{S}{K}^{*}} ( {m}_{K}^{2} ) }  \right\}\eqno(32)
$$
and
$$
A ( {{D}_{S}}^{+}\rightarrow   {\bar{{K}^{0}}}^{*}  {K}  ^{+} ) = {{G}_{F} \over \sqrt{2} } {V}_{cs} {V}_{ud}^{*} {\hat{{a}_{2}}}^{ eff}   {f}_{{K}^{*}} ( 2 {m}_{{K}^{*}} ) \epsilon \cdot {p}_{{D}_{S}} {{F}_{1}}^{{D}_{S}K} ( {m}_{{K}^{*}}^{2} ) \eqno(33)
$$
with
$$
{\hat{{a}_{2}}}^{eff} = {a}_{2} \left\{ 1 + {{\hat{{B}_{0}}}^{ (1) nf} \over  {{F}_{1}}^{{D}_{S}K} ( {m}_{{K}^{*}}^{2} ) }  + {{C}_{1} \over  {a}_{2}} {{\hat{{B}_{0}}}^{ (8) nf} \over  {{F}_{1}}^{{D}_{S}K} ( {m}_{{K}^{*}}^{2} )  }    + {{a}_{1} \over {a}_{2}}{{f}_{{D}_{S}} \over  {f}_{{K}^{*}}} {{\hat{{B}_{0}}}^{ ann} \over  {{F}_{1}}^{{D}_{S}K} ( {m}_{{K}^{*}}^{2} )   }  \right\}\eqno(34)
$$
$ {{B}_{0}}^{ (1) nf}, {{B}_{0}}^{ (8) nf}$ and $ {{B}_{0}}^{ ann}$ are the analogue of $ {{A}_{0}}^{ (1) nf}, {{A}_{0}}^{ (8) nf}$ and  $ {{A}_{0}}^{ ann}$ of Eq. (27). The hatted quantities refer to the channel $  {\bar{{K}^{0}}}^{*}  {K}  ^{+}$.
Using the above equations the amplitude $ {A}^{ ( {K}^{*} K {)}_{S} }$ is written as,
$$
 {A}^{ ( {K}^{*} K {)}_{S} } =   {{G}_{F} \over \sqrt{2} } {V}_{cs} {{V}_{ud}}^{*} { ( 2 {m}_{{K}^{*}} ) \over \sqrt{2} } \left\{  {{a}_{2}}^{ eff}   {f}_{K}  {{A}_{0}}^{{D}_{S}{K}^{*}} ( {m}_{K}^{2} ) +  {\hat{{a}_{2}}}^{ eff}   {f}_{{K}^{*}} {{F}_{1}}^{{D}_{S}K} ( {m}_{{K}^{*}}^{2} )
 \right\}\eqno(35)
$$

The $K$-matrix being real and symmetric is parametrized as follows,
$$
{\bf K} = \pmatrix{
a & b \cr
b & c \cr
}\eqno(36)
$$
On carrying out the unitarization through Eq.(30) one obtains,
$$
{A}^{U} ( {{D}_{S}}^{+}\rightarrow \phi {\pi}^{+} ) =   {{G}_{F} \over \sqrt{2} } {V}_{cs} {V}_{ud}^{*}  {{a}_{1}}^{U,eff} {f}_{\pi}  \epsilon \cdot {p}_{{D}_{S}}   ( 2 {m}_{\phi} ) {{A}_{0}}^{{D}_{S} \phi} ( {m}_{\pi}^{2} ) \eqno(37)
$$
with
$$
{{a}_{1}}^{U,eff} = {{{a}_{1}}^{eff} \over \triangle} \left\{  1  -  i {{k}_{2}}^{3} c
 +  i {{m}_{{K}^{*}} \over \sqrt{2 } {m}_{\phi} } {{k}_{2}}^{3}b F \right\} \eqno(38)
$$
where
$$
F =  {{{a}_{2}}^{eff} \over {{a}_{1}}^{eff}}  {{f}_{K} \over {f}_{\pi}} { {{A}_{0}}^{{D}_{S}{K}^{*}} ( {m}_{K}^{2} ) \over  {{A}_{0}}^{{D}_{S} \phi} ( {m}_{\pi}^{2} )
} +  {{\hat{{a}_{2}}}^{eff} \over {{a}_{1}}^{eff}}  {{f}_{{K}^{*}} \over {f}_{\pi}}  { {{F}_{1}}^{{D}_{S}K} ( {m}_{{K}^{*}}^{2} ) \over {{A}_{0}}^{{D}_{S} \phi} ( {m}_{\pi}^{2} )}
$$
and  $ \triangle= \det{ ({\bf1} - i {{\bf k}}^{3} {\bf K})} $.

If the final-state interactions were elastic, $ b=c=0 $ and  $ \triangle = 1 - i {{k}_{1}}^{3} a $, we would have obtained
$$
{{a}_{1}}^{U, eff}= { {{a}_{1}}^{eff} \over \sqrt{1 + {{k}_{1}}^{6} {a}^{2}} } {e}^{i \delta}\eqno(39)
$$
with $ \delta = tan^{-1} ( a {{k}_{1}}^{3} )$, the elastic P-wave $ \phi \pi $ scattering phase.

In summary, one can always define an effective, and complex $ {a}_{1} $ through the expression in Eq.(37). What we have shown is how effects such as nonfactorized contributions, annihilation and final-state interactions are built into it. A corollary of our point of view is that claims of test of factorization by comparing two-body hadronic rates to semileptonic rates are really nothing more than determinations of $ \mid  {{a}_{1}}^{U, eff}\mid$.

I wish to acknowledge collaboration  with A. B. Santra and F. Ghoddoussi in the research described here and thank the Natural Sciences and Engineering Research Council of Canada for an Individual Research Grant.

\end{document}